\documentclass[pre,aps,floatfix,onecolumn]{revtex4-2}
\usepackage[left=0.9in,right=0.9in,top=0.9in,bottom=0.9in,footskip=.25in]{geometry}
\usepackage{amsmath,amssymb,multirow,epsfig,bm,pifont}
\usepackage{mathtools, float}
\usepackage[toc,page]{appendix}
\usepackage{comment}

\newcommand{\beq}{\begin{equation}}
\newcommand{\eeq}{\end{equation}}
\newcommand{\bea}{\begin{eqnarray}}
\newcommand{\eea}{\end{eqnarray}}

\usepackage[plainpages=false,pdfpagelabels,colorlinks=true,linkcolor=red,urlcolor=blue,citecolor=blue,pdftitle={},pdfauthor={},pdfdisplaydoctitle=true,pdfduplex=DuplexFlipLongEdge]{hyperref}

\begin{document} 

\title{On the Need for Extensible Quantum Compilers with Verification}
\author{Tyler LeBlond}
\affiliation{Computational Sciences and Engineering Division, Oak Ridge National Laboratory}
\author{Xiao Xiao}
\affiliation{Computational Sciences and Engineering Division, Oak Ridge National Laboratory}
\author{Eugene Dumitrescu}
\affiliation{Computational Sciences and Engineering Division, Oak Ridge National Laboratory}
\author{Ryan Bennink}
\affiliation{Computational Sciences and Engineering Division, Oak Ridge National Laboratory}
\author{Alexandru Paler}
\affiliation{Department of Computer Science, Aalto University}

\maketitle

\section{Topic}
\textit{Key-words: Compilation, error correction and mitigation, and codesign and integration.}

In this position paper, we posit that a major Department of Energy (DOE)-funded open-source quantum compilation platform is needed to facilitate: (a) resource optimization at the fault-tolerant layer of the quantum computing software stack, and (b) co-design of that layer of the stack with other layers, and that this platform needs to be extensible and include verification.

\section{Challenge}

The surface code is the most popular quantum error-correcting code (QECC) due to its high error threshold and nearest-neighbor connectivity, which improves its applicability to real hardware~\cite{fowler2012surface}. It is therefore important for the broader community to focus on developing surface code compilation frameworks, and several authors have proposed such frameworks (either abstractly or in software) using so-called lattice surgery, which is the leading technique for implementing fault-tolerant operations with surface codes~\cite{horsman2012surface,fowler2018low,litinski2019game,beverland2022surface,beverland2022assessing,paler2020opensurgery,watkins2023high}. A review of these works reveals that there are multiple layers of complexity to the development of a surface code compilation framework. One must (at least) choose: (a) an input logical circuit instruction set, (b) an output surface code instruction set, (c) a physical layout of surface code patches that includes an appropriate amount of ancillary space for transport and resource state generation, (d) an abstract way to re-express the input logical circuit into surface code operations,  and (e) a scheme for scheduling, routing, and further breaking down the list of abstract surface code instructions into something physically realizable. Because fault-tolerance is expected to increase the overall space-time cost of quantum computation by several orders of magnitude, a fuller optimization of this layer of the stack (b, c, and d) together with its co-design and integration with other layers (a and e) is an important present challenge for the realization of quantum advantage in the fault-tolerant era.


While open-source software packages for surface code compilation have been published, e.g.~\cite{paler2020opensurgery,watkins2023high}, there is no consensus on which compilation strategy will perform the best on practical quantum circuits. 
Moreover, we must ask the question: even if a robust lattice surgery compiler is written, what happens when a more resource-efficient fault-tolerant implementation of logical gates on the surface code is proposed? Despite recent success in implementing the surface code, there is reason to believe that it might not remain the preferred QECC long-term. Its encoding rate does not scale favorably compared with other members of the low-density-parity-check (LDPC) code family (these can encode increasingly many logical qubits with an asymptotically linear number of physical qubits~\cite{breuckmann2021quantum}, which is a very desirable property for resource efficiency reasons). An understanding of how to implement LDPC logical gates fault-tolerantly and resource-efficiently is a subject of ongoing research, though it is suspected that lattice surgery protocols could apply to them. With these realities on the horizon, future quantum compilers should have extensibility to broader classes of QECC in mind.

Additionally, because executed quantum computations have to be both fault-tolerant and correct at the same time, it is critical that the QECC-compiled circuits are verified by the compiler. Several things could go wrong in the process: there might be software bugs in the compiler, the compilation/optimisation methods might be formally incorrect, or the input circuit might even be wrong. However, considering that more general LDPC codes might form the foundation of future large scale computers, it is necessary to devise compilers (or layers of the compilation stack) that can offer guarantees about the compiled circuit. For instance, are the logical gates implemented correctly by the surgery protocols? Does the resulting gate sequence reflect the input circuit? 
Verification can be performed in approximate \cite{paler2018specification} or exact manners \cite{wang2021equivalence}. The latter is exponentially difficult to perform at scale (state vector simulation is very expensive, tensor network approaches are a bit less expensive) while the first offers only bounded guarantees. Nevertheless, approximate verification (i.e., testing) is the only practically viable option.

Because there have been few attempts to create fault-tolerant compilers with verification, doing so represents a research frontier. While verification can be performed at different layers of the quantum computing stack (in bottom-up manner: physical qubit operations, fault-tolerant protocols, logical gates, higher-level logical constructs, functional verification of arithmetic units, \ldots, oracles), efficient and scalable fault-tolerant compilers should at least support verification at the layers of the stack that they touch, i.e., the logical compilation to fault-tolerant operations and the lower-level implementation of those into stabilizer measurement circuits.

\section{Opportunity}

With these considerations in mind, it appears to us that a platform is needed to compare the performance of different quantum compilation strategies on practical circuits and that this platform should be both \textit{extensible} and \textit{verified}. We expect that the level of effort required to develop this platform is significant enough that it is currently only amenable to companies with large, well-funded quantum software divisions. In this setting there is more incentive to produce quantum compilers that are highly tailored toward companies' specific aims, and less incentive to produce an open-source software platform for the community as a whole to explore the vast space of possible choices on the different compilation layers. Therefore, we believe that there is an opportunity for a public entity such as the DOE to fund this platform and drive progress in quantum compilation.

\section{Assessment}


The best overall outcome would be an open-source quantum compilation platform that forms the basis for future compilers written by companies in the private sector. Near-term, it should produce a benchmarking effort that establishes the performance of different surface code compilation frameworks on practical quantum circuits. This can help guide the development and scaling of fault-tolerant quantum computers. Long-term, it should be used as a sandbox for the development of fault-tolerant implementations using different QECC. 

\section{Timelines and Maturity}

Since the advent of noisy intermediate-scale quantum devices, there has been an immense amount of literature published around the topics of QECC and fault-tolerant logical operations using them. This surge has resulted from the need to prepare for a coming fault-tolerant era, which has now been ushered in through initial demonstrations of fault-tolerance by major hardware providers~\cite{google2023suppressing}. 



\bibliography{refs}

\begin{thebibliography}{12}%
\makeatletter
\providecommand \@ifxundefined [1]{%
 \@ifx{#1\undefined}
}%
\providecommand \@ifnum [1]{%
 \ifnum #1\expandafter \@firstoftwo
 \else \expandafter \@secondoftwo
 \fi
}%
\providecommand \@ifx [1]{%
 \ifx #1\expandafter \@firstoftwo
 \else \expandafter \@secondoftwo
 \fi
}%
\providecommand \natexlab [1]{#1}%
\providecommand \enquote  [1]{``#1''}%
\providecommand \bibnamefont  [1]{#1}%
\providecommand \bibfnamefont [1]{#1}%
\providecommand \citenamefont [1]{#1}%
\providecommand \href@noop [0]{\@secondoftwo}%
\providecommand \href [0]{\begingroup \@sanitize@url \@href}%
\providecommand \@href[1]{\@@startlink{#1}\@@href}%
\providecommand \@@href[1]{\endgroup#1\@@endlink}%
\providecommand \@sanitize@url [0]{\catcode `\\12\catcode `\$12\catcode
  `\&12\catcode `\#12\catcode `\^12\catcode `\_12\catcode `\%12\relax}%
\providecommand \@@startlink[1]{}%
\providecommand \@@endlink[0]{}%
\providecommand \url  [0]{\begingroup\@sanitize@url \@url }%
\providecommand \@url [1]{\endgroup\@href {#1}{\urlprefix }}%
\providecommand \urlprefix  [0]{URL }%
\providecommand \Eprint [0]{\href }%
\providecommand \doibase [0]{https://doi.org/}%
\providecommand \selectlanguage [0]{\@gobble}%
\providecommand \bibinfo  [0]{\@secondoftwo}%
\providecommand \bibfield  [0]{\@secondoftwo}%
\providecommand \translation [1]{[#1]}%
\providecommand \BibitemOpen [0]{}%
\providecommand \bibitemStop [0]{}%
\providecommand \bibitemNoStop [0]{.\EOS\space}%
\providecommand \EOS [0]{\spacefactor3000\relax}%
\providecommand \BibitemShut  [1]{\csname bibitem#1\endcsname}%
\let\auto@bib@innerbib\@empty
\bibitem [{\citenamefont {Fowler}\ \emph {et~al.}(2012)\citenamefont {Fowler},
  \citenamefont {Mariantoni}, \citenamefont {Martinis},\ and\ \citenamefont
  {Cleland}}]{fowler2012surface}%
  \BibitemOpen
  \bibfield  {author} {\bibinfo {author} {\bibfnamefont {A.~G.}\ \bibnamefont
  {Fowler}}, \bibinfo {author} {\bibfnamefont {M.}~\bibnamefont {Mariantoni}},
  \bibinfo {author} {\bibfnamefont {J.~M.}\ \bibnamefont {Martinis}},\ and\
  \bibinfo {author} {\bibfnamefont {A.~N.}\ \bibnamefont {Cleland}},\
  }\bibfield  {title} {\bibinfo {title} {Surface codes: Towards practical
  large-scale quantum computation},\ }\href@noop {} {\bibfield  {journal}
  {\bibinfo  {journal} {Physical Review A}\ }\textbf {\bibinfo {volume} {86}},\
  \bibinfo {pages} {032324} (\bibinfo {year} {2012})}\BibitemShut {NoStop}%
\bibitem [{\citenamefont {Horsman}\ \emph {et~al.}(2012)\citenamefont
  {Horsman}, \citenamefont {Fowler}, \citenamefont {Devitt},\ and\
  \citenamefont {Van~Meter}}]{horsman2012surface}%
  \BibitemOpen
  \bibfield  {author} {\bibinfo {author} {\bibfnamefont {C.}~\bibnamefont
  {Horsman}}, \bibinfo {author} {\bibfnamefont {A.~G.}\ \bibnamefont {Fowler}},
  \bibinfo {author} {\bibfnamefont {S.}~\bibnamefont {Devitt}},\ and\ \bibinfo
  {author} {\bibfnamefont {R.}~\bibnamefont {Van~Meter}},\ }\bibfield  {title}
  {\bibinfo {title} {Surface code quantum computing by lattice surgery},\
  }\href@noop {} {\bibfield  {journal} {\bibinfo  {journal} {New Journal of
  Physics}\ }\textbf {\bibinfo {volume} {14}},\ \bibinfo {pages} {123011}
  (\bibinfo {year} {2012})}\BibitemShut {NoStop}%
\bibitem [{\citenamefont {Fowler}\ and\ \citenamefont
  {Gidney}(2018)}]{fowler2018low}%
  \BibitemOpen
  \bibfield  {author} {\bibinfo {author} {\bibfnamefont {A.~G.}\ \bibnamefont
  {Fowler}}\ and\ \bibinfo {author} {\bibfnamefont {C.}~\bibnamefont
  {Gidney}},\ }\bibfield  {title} {\bibinfo {title} {Low overhead quantum
  computation using lattice surgery},\ }\href@noop {} {\bibfield  {journal}
  {\bibinfo  {journal} {arXiv preprint arXiv:1808.06709}\ } (\bibinfo {year}
  {2018})}\BibitemShut {NoStop}%
\bibitem [{\citenamefont {Litinski}(2019)}]{litinski2019game}%
  \BibitemOpen
  \bibfield  {author} {\bibinfo {author} {\bibfnamefont {D.}~\bibnamefont
  {Litinski}},\ }\bibfield  {title} {\bibinfo {title} {A game of surface codes:
  Large-scale quantum computing with lattice surgery},\ }\href@noop {}
  {\bibfield  {journal} {\bibinfo  {journal} {Quantum}\ }\textbf {\bibinfo
  {volume} {3}},\ \bibinfo {pages} {128} (\bibinfo {year} {2019})}\BibitemShut
  {NoStop}%
\bibitem [{\citenamefont {Beverland}\ \emph
  {et~al.}(2022{\natexlab{a}})\citenamefont {Beverland}, \citenamefont
  {Kliuchnikov},\ and\ \citenamefont {Schoute}}]{beverland2022surface}%
  \BibitemOpen
  \bibfield  {author} {\bibinfo {author} {\bibfnamefont {M.}~\bibnamefont
  {Beverland}}, \bibinfo {author} {\bibfnamefont {V.}~\bibnamefont
  {Kliuchnikov}},\ and\ \bibinfo {author} {\bibfnamefont {E.}~\bibnamefont
  {Schoute}},\ }\bibfield  {title} {\bibinfo {title} {Surface code compilation
  via edge-disjoint paths},\ }\href@noop {} {\bibfield  {journal} {\bibinfo
  {journal} {PRX Quantum}\ }\textbf {\bibinfo {volume} {3}},\ \bibinfo {pages}
  {020342} (\bibinfo {year} {2022}{\natexlab{a}})}\BibitemShut {NoStop}%
\bibitem [{\citenamefont {Beverland}\ \emph
  {et~al.}(2022{\natexlab{b}})\citenamefont {Beverland}, \citenamefont
  {Murali}, \citenamefont {Troyer}, \citenamefont {Svore}, \citenamefont
  {Hoeffler}, \citenamefont {Kliuchnikov}, \citenamefont {Low}, \citenamefont
  {Soeken}, \citenamefont {Sundaram},\ and\ \citenamefont
  {Vaschillo}}]{beverland2022assessing}%
  \BibitemOpen
  \bibfield  {author} {\bibinfo {author} {\bibfnamefont {M.~E.}\ \bibnamefont
  {Beverland}}, \bibinfo {author} {\bibfnamefont {P.}~\bibnamefont {Murali}},
  \bibinfo {author} {\bibfnamefont {M.}~\bibnamefont {Troyer}}, \bibinfo
  {author} {\bibfnamefont {K.~M.}\ \bibnamefont {Svore}}, \bibinfo {author}
  {\bibfnamefont {T.}~\bibnamefont {Hoeffler}}, \bibinfo {author}
  {\bibfnamefont {V.}~\bibnamefont {Kliuchnikov}}, \bibinfo {author}
  {\bibfnamefont {G.~H.}\ \bibnamefont {Low}}, \bibinfo {author} {\bibfnamefont
  {M.}~\bibnamefont {Soeken}}, \bibinfo {author} {\bibfnamefont
  {A.}~\bibnamefont {Sundaram}},\ and\ \bibinfo {author} {\bibfnamefont
  {A.}~\bibnamefont {Vaschillo}},\ }\bibfield  {title} {\bibinfo {title}
  {Assessing requirements to practical quantum advantage},\ }\href@noop {}
  {\bibfield  {journal} {\bibinfo  {journal} {arXiv preprint arXiv:2211.07629}\
  } (\bibinfo {year} {2022}{\natexlab{b}})}\BibitemShut {NoStop}%
\bibitem [{\citenamefont {Paler}\ and\ \citenamefont
  {Fowler}(2020)}]{paler2020opensurgery}%
  \BibitemOpen
  \bibfield  {author} {\bibinfo {author} {\bibfnamefont {A.}~\bibnamefont
  {Paler}}\ and\ \bibinfo {author} {\bibfnamefont {A.~G.}\ \bibnamefont
  {Fowler}},\ }\bibfield  {title} {\bibinfo {title} {Opensurgery for
  topological assemblies},\ }in\ \href@noop {} {\emph {\bibinfo {booktitle}
  {2020 IEEE Globecom Workshops (GC Wkshps}}}\ (\bibinfo {organization}
  {IEEE},\ \bibinfo {year} {2020})\ pp.\ \bibinfo {pages} {1--4}\BibitemShut
  {NoStop}%
\bibitem [{\citenamefont {Watkins}\ \emph {et~al.}(2023)\citenamefont
  {Watkins}, \citenamefont {Nguyen}, \citenamefont {Seshadri}, \citenamefont
  {Watkins}, \citenamefont {Pearce}, \citenamefont {Lau},\ and\ \citenamefont
  {Paler}}]{watkins2023high}%
  \BibitemOpen
  \bibfield  {author} {\bibinfo {author} {\bibfnamefont {G.}~\bibnamefont
  {Watkins}}, \bibinfo {author} {\bibfnamefont {H.~M.}\ \bibnamefont {Nguyen}},
  \bibinfo {author} {\bibfnamefont {V.}~\bibnamefont {Seshadri}}, \bibinfo
  {author} {\bibfnamefont {K.}~\bibnamefont {Watkins}}, \bibinfo {author}
  {\bibfnamefont {S.}~\bibnamefont {Pearce}}, \bibinfo {author} {\bibfnamefont
  {H.-K.}\ \bibnamefont {Lau}},\ and\ \bibinfo {author} {\bibfnamefont
  {A.}~\bibnamefont {Paler}},\ }\bibfield  {title} {\bibinfo {title} {A high
  performance compiler for very large scale surface code computations},\
  }\href@noop {} {\bibfield  {journal} {\bibinfo  {journal} {arXiv preprint
  arXiv:2302.02459}\ } (\bibinfo {year} {2023})}\BibitemShut {NoStop}%
\bibitem [{\citenamefont {Breuckmann}\ and\ \citenamefont
  {Eberhardt}(2021)}]{breuckmann2021quantum}%
  \BibitemOpen
  \bibfield  {author} {\bibinfo {author} {\bibfnamefont {N.~P.}\ \bibnamefont
  {Breuckmann}}\ and\ \bibinfo {author} {\bibfnamefont {J.~N.}\ \bibnamefont
  {Eberhardt}},\ }\bibfield  {title} {\bibinfo {title} {Quantum low-density
  parity-check codes},\ }\href@noop {} {\bibfield  {journal} {\bibinfo
  {journal} {PRX Quantum}\ }\textbf {\bibinfo {volume} {2}},\ \bibinfo {pages}
  {040101} (\bibinfo {year} {2021})}\BibitemShut {NoStop}%
\bibitem [{\citenamefont {Paler}\ and\ \citenamefont
  {Devitt}(2018)}]{paler2018specification}%
  \BibitemOpen
  \bibfield  {author} {\bibinfo {author} {\bibfnamefont {A.}~\bibnamefont
  {Paler}}\ and\ \bibinfo {author} {\bibfnamefont {S.~J.}\ \bibnamefont
  {Devitt}},\ }\bibfield  {title} {\bibinfo {title} {Specification format and a
  verification method of fault-tolerant quantum circuits},\ }\href@noop {}
  {\bibfield  {journal} {\bibinfo  {journal} {Physical Review A}\ }\textbf
  {\bibinfo {volume} {98}},\ \bibinfo {pages} {022302} (\bibinfo {year}
  {2018})}\BibitemShut {NoStop}%
\bibitem [{\citenamefont {Wang}\ \emph {et~al.}(2021)\citenamefont {Wang},
  \citenamefont {Li},\ and\ \citenamefont {Ying}}]{wang2021equivalence}%
  \BibitemOpen
  \bibfield  {author} {\bibinfo {author} {\bibfnamefont {Q.}~\bibnamefont
  {Wang}}, \bibinfo {author} {\bibfnamefont {R.}~\bibnamefont {Li}},\ and\
  \bibinfo {author} {\bibfnamefont {M.}~\bibnamefont {Ying}},\ }\bibfield
  {title} {\bibinfo {title} {Equivalence checking of sequential quantum
  circuits},\ }\href@noop {} {\bibfield  {journal} {\bibinfo  {journal} {IEEE
  Transactions on Computer-Aided Design of Integrated Circuits and Systems}\
  }\textbf {\bibinfo {volume} {41}},\ \bibinfo {pages} {3143} (\bibinfo {year}
  {2021})}\BibitemShut {NoStop}%
\bibitem [{goo(2023)}]{google2023suppressing}%
  \BibitemOpen
  \bibfield  {title} {\bibinfo {title} {Suppressing quantum errors by scaling a
  surface code logical qubit},\ }\href@noop {} {\bibfield  {journal} {\bibinfo
  {journal} {Nature}\ }\textbf {\bibinfo {volume} {614}},\ \bibinfo {pages}
  {676} (\bibinfo {year} {2023})}\BibitemShut {NoStop}%
\end{thebibliography}%

\appendix
This manuscript has been authored by UT-Battelle, LLC, under contract DE-AC05-00OR22725 with the US Department of Energy (DOE). The US government retains and the publisher, by accepting the article for publication, acknowledges that the US government retains a nonexclusive, paid-up, irrevocable, worldwide license to publish or reproduce the published form of this manuscript, or allow others to do so, for US government purposes. DOE will provide public access to these results of federally sponsored research in accordance with the DOE Public Access Plan (https://www.energy.gov/downloads/doe-public-access-plan).
\end{document}